\documentclass[11pt,preprint,prd,aps,nofootinbib,showpacs,a4paper]{revtex4}
\usepackage[dvips]{graphicx}
\usepackage{amsfonts, color, float, bbm, amsmath}
\usepackage{amssymb}
\usepackage{pstricks}

\def\beq {\begin{equation}}
\def\eeq {\end{equation}}
\def\bea {\begin{eqnarray}}
\def\eea {\end{eqnarray}}

\def\A {\mathcal{A}}

\def\DKpp{D^+  \to K^- \pi^+\pi^+ }

\begin{document}

\preprint{UAB--FT--668}

\title{$K\pi$ form factors and final state interactions in $D^+\to K^-\pi^+\pi^+$ decays}

\author{D.~R.~Boito}
\email{boito@ifae.es}
\author{R.~Escribano}
\email{escribano@ifae.es}
\affiliation{Grup de F\'{\i}sica Te\`orica and IFAE, Universitat Aut\`onoma de Barcelona,
E-08193 Bellaterra (Barcelona), Spain}
%\date{\today}

\begin{abstract}
We present a model for the decay $D^+\to K^-\pi^+\pi^+$.  The weak interaction
part of this reaction is described using the effective weak
Hamiltonian in the factorisation approach.  Hadronic final state
interactions are taken into account through the $K\pi$  scalar and vector
form factors fulfilling analyticity, unitarity and chiral symmetry
constraints.  The model has only two free parameters that are fixed
from experimental branching ratios.  We show that the modulus and phase of the
$S$ wave thus obtained agree nicely with experiment up to 1.55~GeV.
We perform Monte Carlo simulations to compare the predicted Dalitz
plot with experimental analyses.  Allowing for a global phase
difference between the $S$ and $P$ waves of $-65^\circ$, the Dalitz plot of the
$D^+\to K^-\pi^+\pi^+$ decay, the $K\pi$ invariant mass spectra and
the total branching ratio due to $S$-wave interactions are well
reproduced.
\end{abstract}


% PACS, the Physics and Astronomy Classification Scheme.
%\keywords{Suggested keywords}%Use showkeys class option if keyword display desired
\maketitle

\section{Introduction}
\label{intro}

%%Experimental facts
In 2002, the analysis of $\DKpp$ decays performed by the E791
collaboration revealed that approximately 50\% of these decays
proceed through a low-mass scalar resonance with isospin $1/2$: the
$K_0^*(800)$, also called the $\kappa$~\cite{Aitala:2002kr}. As a
matter of fact, the $\kappa$ was the second elusive scalar to be
firmly detected in $D^+$ decays since the scalar-isoscalar $f_0(600)$,
or $\sigma$, had been detected by the same collaboration in $D^+\to
\pi^+ \pi^- \pi^+$ \cite{E791sigma}. More recently, the $\DKpp$ decay
was revisited by E791~\cite{Aitala:2005yh} and two other experiments
produced analyses based on larger data samples, namely
FOCUS~\cite{Pennington:2007se,Focus2009} and
CLEO~\cite{Bonvicini:2008jw}.  The main conclusions of the pioneering
E791 work have been confirmed in both cases.

%%The kappa pole in scattering
In the past, many analyses of $K\pi$ scattering data had already
claimed the presence of the $\kappa$ pole in the scattering amplitude
\cite{kappapoles1,kappapoles2,kappapoles3,JOPscatt}. The most precise and model independent
determination of its position in the second Riemann sheet was produced
in Ref.~\cite{MoussallamKappa}, following the method put forward for
the $\sigma$ in Ref.~\cite{CCL}.  Using Roy's equations for $K\pi$
scattering \cite{Roy} and Chiral Perturbation Theory
(ChPT)~\cite{ChPT} Descotes-Genon and Moussallam found $m_\kappa =
658\pm 13$ MeV and $\Gamma_\kappa = 557\pm 24$~MeV~\cite{MoussallamKappa}.

%% Difficulties, problems and Oller's work
Although the experimental results are sound and the $\kappa$ pole is
at present theoretically well known, a comprehensive and successful
description of the reaction $\DKpp$ is still not available (for a recent
review see Ref.~\cite{Reviews}). Experimentalists, for the want of a
better framework, commonly fit their data with the isobar model which
consists of a weighted sum of Breit-Wigner-like propagators. Often, a
complex constant is added to the amplitude in order to account for the
non-resonant decays.  It is known, nevertheless, that the adoption of
Breit-Wigner functions to describe the effect of scalar resonances is
problematic. Some of the deficiencies of this approach are discussed
in Ref.~\cite{Oller:2004xm} where Oller proposed the substitution of these
functions in the $S$ wave by expressions based on unitarised
ChPT~\cite{UniChPT}. This model provides a good description of the
data but, since the weak part of the decay was not tackled, the
relative weight of the amplitudes remain  arbitrary complex
parameters to be determined from the fit.

%%Weak decay
Little progress has been achieved in the treatment of weak decays of
charmed mesons since the seminal papers by Bauer, Stech and
Wirbel~\cite{Wirbel:1985ji,Bauer:1986bm}. This fact stems from the
mass of the $c$-quark that lies between the heavy and the light
domains, rendering heavy-quark approaches or the use of chiral
symmetry less trustworthy.  A first attempt to describe the decay
$\DKpp$ from first principles was made by Diakonou and Diakonos in
Ref.~\cite{Diakonou:1989sf}. In their work, the weak amplitude was
described within na\"ive factorisation with the weak Hamiltonian of
Refs.~\cite{Wirbel:1985ji,Bauer:1986bm} and the final state
interactions (FSIs) were implemented by means of Breit-Wigner type
$K\pi$ form factors. They considered the contribution of two
resonances, namely the $K^*(892)$ and the $K_0^*(1430)$. In the
light of the present empirical data it is clear that this model cannot
provide a good description of the decay.  In
Ref.~\cite{Diakonou:1989sf}, the decay is mainly driven by the 
$K^*(892)$ whereas the analyses of
Refs.~\cite{Aitala:2002kr,Aitala:2005yh, Pennington:2007se,Focus2009,
  Bonvicini:2008jw} show that the decay is largely dominated by $K\pi$
pairs in an $S$-wave state. On average, the total scalar signal
amounts to 82\%~\cite{Amsler:2008zzb}.  Hence, a more comprehensive
model for the whole scalar contribution is needed to provide a good
description of the data.  A first step in this direction was taken in
Refs.~\cite{Gardner:2001gc,Gardner2} where the $\pi\pi$ scalar signal
in $B\to\pi\pi\pi$ decays was considered. In this framework,
factorisation is assumed for the weak amplitude and the $\pi\pi$
scalar form factor, constrained by chiral dynamics and unitarity,
provides the description of~FSIs~\cite{Meissner:2000bc}.  In
Refs.~\cite{Furman:2005xp,ElBennich:2006yi}, a similar description was
utilised to describe the $S$ wave in $B\to\pi\pi K$ and $B\to K\bar K
K$ decays.  Using the same method, $S$-wave FSIs have also been
considered in the decay $D^+\to \pi^+ \pi^-
\pi^+$~\cite{Boito:2008zk}.  More recently, $K\pi$ form factors have
been employed in the description of FSIs in $B^\pm\to
K^\pm\pi^\mp\pi^\pm$ decays~\cite{BenoitBKpipi}. In the
present work, we follow the same general scheme where a factorised
weak decay amplitude is dressed with FSIs by means of non-perturbative
$K\pi$ form factors.

 %%This work

 For the weak vertex, we employ the effective weak Hamiltonian of
 Refs.~\cite{Wirbel:1985ji,Bauer:1986bm} within na\"ive
 factorisation. Although the assumption of factorisation is less
 reliable for the $c$-quark mass scale, it has been successfully
 applied to $D$ decays in several recent papers~\cite{Boito:2008zk,
   Bediaga:1996ue, Bediagaetal2, Rosenfeld,Cheng:2002ai, f0}.
 However, one should consider the Wilson coefficients as
 phenomenological parameters to compensate for the deficiencies of
 factorisation~\cite{Buras:1994ij}.  The phenomenological values are close to the calculated
 ones~\cite{Heff}  but  have larger errors than in applications to $B$ decays.  The
 weak amplitude thus obtained receives contributions from
 colour-allowed and colour-suppressed topologies. In the latter, the
 $K\pi$ form factors appear manifestly and the construction of the
 final state is straightforward.  The colour-allowed topology is more
 involved but, assuming the decay to be mediated by resonances as
 suggested by the experimental results, the FSIs in this case can also
 be written in terms of $K\pi$ form
 factors~\cite{Gardner:2001gc,Boito:2008zk}.  Therefore, in our
 description the hadronic FSIs are fully taken into account by the
 $K\pi$ scalar and vector form factors.

%Kpi form factors 

 Both form factors have received attention in recent years and are now
 well known in the energy regime relevant to $\DKpp$ decays. The scalar
 component was studied in a framework that incorporates all the known
 theoretical constraints in
 Refs.~\cite{Jamin:2001zq,JOP2,JOP3}. Analyticity, unitarity, chiral
 symmetry, the large-$N_c$ limit of QCD, and the coupling to $K\eta$
 and $K\eta'$ channels were taken into account. The results were
 subsequently updated and we employ in this work the state-of-the-art
 version given in Ref.~\cite{Jamin:2006tj}.  The vector form factor,
 in its turn, can be studied in $\tau^- \to K \pi \nu_\tau$ decays~\cite{Jamin:2006tk,JPP2,Moussallam,Boito:2008fq}, where
 the kinematical range is very similar to the one considered in this
 paper.  A prediction for this form factor within Resonance Chiral
 Theory (RChT)~\cite{RChT} was presented in Ref.~\cite{Jamin:2006tk}
 and, after the appearance of the detailed spectrum measured by the
 Belle collaboration~\cite{Belle}, a fit was performed in
 Ref.~\cite{JPP2}. Here we employ a slightly different description
 which fulfils analyticity constraints and that was successfully
 fitted to the Belle spectrum in Ref.~\cite{Boito:2008fq}.

%%Organisation of the paper
Our paper is organised as follows. In Section~\ref{amplitudes} we present our model
and discuss previous treatments of the same decay found in the
literature.  The numerical results are worked out in
Section~\ref{results}. Finally, we give a summary and discuss the results in
Section~\ref{summary}. Details about the construction of the $K\pi$
form factors employed in this work are relegated to the Appendix.

%%%% SECTION II: THE MODEL %%%%%%%%%%%%%%%%%%%%%%%%%%%%%%%%%
\section{Theoretical framework}
\label{amplitudes}

Our phenomenological description of the weak process $D^+\to K^-\pi^+\pi^+$ is based on the
effective Hamiltonian
\begin{equation}
\label{effH}
{\cal H}_{\rm eff}=\frac{G_F}{\sqrt{2}}V_{cs}V_{ud}^\ast
[C_1(\mu)O_1+C_2(\mu)O_2]+\mbox{h.c.}\ ,
\end{equation}
where $G_F=1.16637\times 10^{-5}\ \mbox{GeV}^{-2}$ is the Fermi decay constant
\cite{Amsler:2008zzb}, $V_{cs}V_{ud}^\ast=1-\lambda^2$, in the Wolfenstein parametrisation \cite{Wolfenstein:1983yz} with $\lambda\equiv\sin\theta_C=0.2257$ \cite{Amsler:2008zzb},
$C_{1,2}(\mu)$ are short distance Wilson coefficients computed at the renormalisation scale
$\mu={\cal O}(m_c)$, and $O_{1,2}$ are the local four-quark operators
\begin{equation}
\label{effO12}
\begin{array}{c}
O_1=[\bar c_i\gamma^\mu (1-\gamma_5) s_i][\bar d_j\gamma_\mu (1-\gamma_5) u_j]\ ,\\
O_2=[\bar c_i\gamma^\mu (1-\gamma_5) s_j][\bar d_j\gamma_\mu (1-\gamma_5) u_i]\ ,
\end{array}
\end{equation}
with $(i,j=1,2,3)$ denoting colour indices.
At the quark level, the decay $D^+\to K^-\pi^+\pi^+$ is driven by the transition $c\to s u\bar d$,
\textit{i.e.}~four different quark flavours are involved.
In this case, only the two tree operators in Eq.~(\ref{effO12}) have to be taken into account.

The amplitude for $D^+\to K^-\pi^+\pi^+$ is given by the matrix element
$\langle K^-\pi^+\pi^+|{\cal H}_{\rm eff}|D^+\rangle$.
We assume the factorisation approach to hold at leading order
(in $\Lambda_{\rm QCD}/m_c$ and $\alpha_s$) and as a consequence
the amplitude is written in terms of colour allowed and suppressed contributions,
${\cal A}_1$ and ${\cal A}_2$ respectively, as
\begin{equation}
\label{amplitude}
{\setlength\arraycolsep{2pt} 
\begin{array}{rcl}
{\cal A}(D^+\to K^-\pi^+\pi^+)&=&
\displaystyle
\frac{G_F}{\sqrt{2}}\cos^2\theta_C(a_1 {\cal A}_1+a_2 {\cal A}_2)
+(\pi^+_1\leftrightarrow\pi^+_2)\\[2ex]
&=&
\displaystyle
\frac{G_F}{\sqrt{2}}\cos^2\theta_C
[a_1\langle K^-\pi^+_1|\bar s\gamma^\mu(1-\gamma_5)c|D^+\rangle
        \langle\pi^+_2|\bar u\gamma_\mu(1-\gamma_5)d|0\rangle\\[2ex]
&&
+a_2\langle K^-\pi^+_1|\bar s\gamma^\mu(1-\gamma_5)d|0\rangle
        \langle\pi^+_2|\bar u\gamma_\mu(1-\gamma_5)c|D^+\rangle]
%\\[1ex]
%&&
+(\pi^+_1\leftrightarrow\pi^+_2)\ ,
\end{array}
}
\end{equation}
where the last term accounts for the presence of two identical pions in the final state. 
The QCD factors $a_{1,2}(\mu)$ are related to $C_{1,2}(\mu)$ as follows:
\begin{equation}
\label{a12}
a_1(\mu)=C_1(\mu)+\frac{1}{N_c}C_2(\mu)\ ,\quad
a_2(\mu)=C_2(\mu)+\frac{1}{N_c}C_1(\mu)\ ,
\end{equation}
where $N_c=3$ is the number of colours.
For these factors we use the phenomenological values
\begin{equation}
\label{a12pheno}
a_1=1.2\pm 0.1\ ,\qquad a_2=-0.5\pm 0.1\ ,
\end{equation}
obtained from different analyses of two-body $D$ meson decays \cite{Buras:1994ij}.

The non-perturbative hadronic matrix elements in Eq.~(\ref{amplitude})
involve several Lorentz invariant form factors.
We first consider those related to the ${\cal A}_2$ contribution.
The transition $D^+\to K^-\pi^+$ appearing in ${\cal A}_1$ is more involved and requires a separate analysis.
The matrix element from the vacuum to the $K\pi$ final state is given by
\begin{equation}
\label{Kpiff}
\langle K^-\pi^+_1|\bar s\gamma^\mu d|0\rangle=
\left[(p_K-p_{\pi_1})^\mu-\frac{m_K^2-m_\pi^2}{q^2}q^\mu\right]F_+^{K\pi}(q^2)
+\frac{m_K^2-m_\pi^2}{q^2}q^\mu F_0^{K\pi}(q^2)\ ,
\end{equation}
where $q=p_K+p_{\pi_1}$ and $F_{+,0}^{K\pi}(q^2)$ are the $K\pi$ vector and scalar form factors.
Analogously, the transition $D^+\to\pi^+$ is given by
\begin{equation}
\label{Dpiff}
\langle \pi^+_2|\bar u\gamma^\mu c|D^+\rangle=
\left[(p_D+p_{\pi_2})^\mu-\frac{m_D^2-m_\pi^2}{q^2}q^\mu\right]F_+^{D\pi}(q^2)
+\frac{m_D^2-m_\pi^2}{q^2}q^\mu F_0^{D\pi}(q^2)\ ,
\end{equation}
where now $q=p_D-p_{\pi_2}$ and $F_{+,0}^{D\pi}(q^2)$ are the $D\pi$ vector and scalar transition form factors, respectively.
The amplitude ${\cal A}_2$ then reads
\begin{equation}
\label{A2}
{\setlength\arraycolsep{2pt}
\begin{array}{rcl}
{\cal A}_2
&=&
\left[m_{K\pi_2}^2-m_{\pi_1\pi_2}^2-
\displaystyle\frac{(m_K^2-m_\pi^2)(m_D^2-m_\pi^2)}{m_{K\pi_1}^2}\right]
F_+^{K\pi}(m_{K\pi_1}^2)F_+^{D\pi}(m_{K\pi_1}^2)\\[2ex]
&&
+\displaystyle\frac{(m_K^2-m_\pi^2)(m_D^2-m_\pi^2)}{m_{K\pi_1}^2}
F_0^{K\pi}(m_{K\pi_1}^2)F_0^{D\pi}(m_{K\pi_1}^2)\ ,
\end{array}
}
\end{equation}
where the Mandelstam variables are defined as
\begin{equation}
\label{Mandelstam}
m_{K\pi_1}^2\equiv (p_K+p_{\pi_1})^2\ , \qquad
m_{K\pi_2}^2\equiv (p_K+p_{\pi_2})^2\ , \qquad
m_{\pi_1\pi_2}^2\equiv (p_{\pi_1}+p_{\pi_2})^2\ ,
\end{equation}
with $m_{K\pi_1}^2+m_{K\pi_2}^2+m_{\pi_1\pi_2}^2=m_D^2+m_K^2+2m_\pi^2$.

In our analysis, we use a simple pole prescription for the $D\pi$ transition form factors, 
\begin{equation}
\label{DpiffBelle}
F_{+,0}^{D\pi}(q^2)=\frac{F_{+,0}^{D\pi}(0)}{1-q^2/m_{\rm pole}^2}\ ,
\end{equation}
with $m_{\rm pole}=m_{D^{\ast 0}}$ for the vector case and
$m_{\rm pole}=m_{D^{\ast 0}_0}$ for the scalar one.
The normalisation constant is by construction the same in both cases
$F_+^{D\pi}(0)=F_0^{D\pi}(0)$.
This parametrisation agrees with experiment.
The analysis performed by the Belle Coll.~on $D^0\to\pi^- l^+\nu$ data
gives for the simple pole model $m_{\rm pole}(1^{--})=1.97\pm 0.09$ \cite{Widhalm:2006wz},
which is compatible with the PDG value $m_{D^{\ast\pm}}=2.01\ \mbox{GeV}$
\cite{Amsler:2008zzb}.
Then, in Eq.~(\ref{DpiffBelle}) we take $F_+^{D\pi}(0)=0.624$ from Ref.~\cite{Widhalm:2006wz}
and $m_{D^{\ast 0}}=2007\ \mbox{MeV}$ and $m_{D^{\ast 0}_0}=2.352\pm 0.050\ \mbox{GeV}$
from Ref.~\cite{Amsler:2008zzb}.

For the $K\pi$ vector and scalar form factors, we employ the same expressions that were used in the successful reanalysis of $\tau^-\to K \pi\nu_\tau$ decays performed in Ref.~\cite{Boito:2008fq}.
Since the kinematical region for the $K\pi$ system available in $D\to K\pi\pi$ decays,
$m_K+m_\pi\leq m_{K\pi}\leq m_D-m_\pi$, is very similar to that of $\tau^-\to K \pi\nu_\tau$ decays,
$m_K+m_\pi\leq m_{K\pi}\leq m_\tau$, we consider this choice appropriate.
Both form factors are constructed such that they fulfil constraints posed by analyticity and unitarity.
Because of these properties, the form factors satisfy an $n$-subtracted dispersion relation,
which in the elastic region admit the well-known Omn\`es solution
\cite{Omnes:1958hv}.
For the $K\pi$ vector form factor $F_+^{K\pi}(s)$,
a good description of the experimental measurement of $\tau^-\to K \pi\nu_\tau$
was achieved by incorporating two vector resonances and working with a three-times-subtracted dispersion relation in order to suppress higher-energy contributions \cite{Boito:2008fq}.
The additionally required scalar $K\pi$ form factor $F_0^{K\pi}(s)$
had been calculated in the  framework of RChT 
and solving dispersion relations for a three-body coupled-channel problem in Ref.~\cite{Jamin:2001zq}.
Here, we use the recent numerical update of Ref.~\cite{Jamin:2006tj}.
The details of the form factors used in this work can be found in Appendix \ref{formfactors}.

Now, we turn our attention to the form factors associated with the ${\cal A}_1$ contribution.
The form factor denoting the transition from the vacuum to a pion final state is nothing else than
\begin{equation}
\label{piondc}
\langle\pi^+_2|\bar u\gamma_\mu(1-\gamma_5)d|0\rangle=i f p_{\pi_2}\ ,
\end{equation}
where the constant $f$ equals at lowest order in the chiral expansion the pion decay constant
$f=f_\pi=\sqrt{2}F_\pi=130.5$ MeV.
The form factors related to the transition $D^+\to K^-\pi^+$ are more complicated.
On general grounds, the matrix element 
$\langle K^-\pi^+_1|\bar s\gamma^\mu(1-\gamma_5)c|D^+\rangle$
can be written in terms of four different form factors \cite{Kuhn:1992nz}.
But, when saturated with
$\langle\pi^+_2|\bar u\gamma_\mu(1-\gamma_5)d|0\rangle$
only one of those form factors survives, $F_4$, and the amplitude ${\cal A}_1$ becomes
\begin{equation}
\label{A1}
{\cal A}_1=
-i f_\pi m_\pi^2 F_4(m_{K\pi_1}^2,m_{K\pi_2}^2)\ .
\end{equation}
Since this amplitude is proportional to $m_\pi^2$ one would expect it is negligible,
as presumed in Ref.~\cite{Bediaga:1996ue}.
If this were the case, however, the decay $D^+\to K^-\pi^+\pi^+$ would be dominated by the
$P$-wave contribution (as demonstrated in Table~\ref{tab3} of Section~\ref{results})
in contradiction with experiment~\cite{Amsler:2008zzb}.
This fact forces one to consider the ${\cal A}_1$ contribution in detail.
Unfortunately, the contribution of $F_4$ to semileptonic decays,
$D^+\to K^-\pi^+l^+\nu_l$ ($l=e, \mu$), is proportional to the lepton masses and neglected~\cite{Bajc:1997nx}.
Consequently, one has to resort to theoretical models.

Several methods have been considered in the literature.
Most of them are based on the assumption that the $D^+\to K^-\pi^+$ transition is driven by intermediate resonances, mainly vectors and scalars in this case.
We will not take into account the contribution of tensor resonances.
%Experimentally, their effects amount to less than 1\% of the total signal.
In the simplest case, one can consider the exchange of a single vector and scalar resonance using a Breit-Wigner parametrisation.
For instance, in the paper by Diakonou and Diakonos \cite{Diakonou:1989sf}
the colour allowed contribution is written via the exchange of
$K^\ast(892)$ and $K^\ast_0(1430)$ resonances as
\begin{equation}
\label{A1Diakonos}
{\setlength\arraycolsep{2pt}
\begin{array}{rcl}
{\cal A}_1 &=&
\left[\displaystyle\sum_{\rm pol}
\frac{\langle K^-\pi_1^+|\bar K^\ast\rangle
         \langle\bar K^\ast|\bar s\gamma^\mu(1-\gamma_5)c|D^+\rangle}
{m_{\bar K^\ast}^2-m_{K\pi_1}^2}
+\frac{\langle K^-\pi_1^+|\bar K^\ast_0\rangle
           \langle\bar K^\ast_0|\bar s\gamma^\mu(1-\gamma_5)c|D^+\rangle}
{m_{\bar K^\ast_0}^2-m_{K\pi_1}^2}\right]\\[2ex]
&& \times\langle\pi^+_2|\bar u\gamma_\mu(1-\gamma_5)d|0\rangle\ ,
\end{array}
}
\end{equation}
while the colour suppressed contribution is given by Eq.~(\ref{A2})
but with monopole $K\pi$ form factors,
$F_{+,0}^{K\pi}(q^2)=F_{+,0}^{K\pi}(0)/(1-q^2/m_{\rm pole}^2)$,
with $m_{\rm pole}=m_{\bar K^\ast(892)}$ for the vector and
$m_{\rm pole}=m_{\bar K^\ast_0(1430)}$ for the scalar.
Taking the matrix elements from Refs.~\cite{Wirbel:1985ji,Bauer:1986bm} one gets
\begin{equation}
\label{A1Diakonosv2}
{\cal A}_1=
\frac{f_\pi g_{\bar K^\ast K\pi}m_{\bar K^\ast}N(m_{\bar K^\ast}^2)F_+^{D\bar K^\ast}(m_\pi^2)}
        {m_{\bar K^\ast}^2-m_{K\pi_1}^2-i m_{\bar K^\ast}\Gamma_{\bar K^\ast}}+
\frac{f_\pi g_{\bar K^\ast_0 K\pi}m_{\bar K^\ast_0}
         (m_D^2-m_{\bar K^\ast_0}^2)F_0^{D\bar K^\ast_0}(m_\pi^2)}
        {m_{\bar K^\ast_0}^2-m_{K\pi_1}^2-i m_{\bar K^\ast_0}\Gamma_{\bar K^\ast_0}}\ ,
\end{equation}
where $N(q^2)=m_D^2+m_K^2+2m_\pi^2-2m_{\pi_1\pi_2}^2-q^2-M(q^2)$,
$M(q^2)=(m_K^2-m_\pi^2)(m_D^2-m_\pi^2)/q^2$,
$g_{\bar K^\ast K\pi}(g_{\bar K^\ast_0 K\pi})$ are dimensionless couplings associated to
$\langle K^-\pi_1^+|\bar K^\ast (\bar K^\ast_0)\rangle$, and
$F_+^{D\bar K^\ast}(m_\pi^2)$ and $F_0^{D\bar K^\ast_0}(m_\pi^2)$
are pertinent vector and scalar transition form factors evaluated at $q^2=m_\pi^2$.
Again, a monopole form is assumed,
\begin{equation}
\label{DVDSff}
F_+^{D\bar K^\ast}(q^2)=\frac{F_+^{D\bar K^\ast}(0)}{1-q^2/m_{\rm pole}^2}\ ,
\qquad
F_0^{D\bar K^\ast_0}(q^2)=\frac{F_0^{D\bar K^\ast_0}(0)}{1-q^2/m_{\rm pole}^2}\ ,
\end{equation}
with $m_{\rm pole}=m_{D_s^\pm}$ in both cases \cite{Wirbel:1985ji,Bauer:1986bm}.

Experimental data collected in Tables~\ref{tab1} and~\ref{tab2} indicate that the vector contribution to the total signal is largely dominated by the exchange of $K^\ast(892)$.
Hence, a Breit-Wigner parametrisation with a single vector resonance,
as considered in Ref.~\cite{Diakonou:1989sf},
should be a reasonable approximation to the vector induced signal.
This is not the case for the scalar one, where the contribution of $K^\ast_0(1430)$ is marginal.
Besides, the possible $K^\ast_0(800)$ or $\kappa$ and non-resonant contributions are not accounted for in Eq.~(\ref{A1Diakonosv2}).
Therefore, a more elaborated prescription taking into account the whole scalar contribution is mandatory.
Here, we follow Ref.~\cite{Gardner:2001gc}
and write the colour allowed amplitude ${\cal A}_1$ in terms of the scalar and vector $K\pi$ form factors.
We briefly summarise the method applied to our case.
The $D^+\to K^-\pi^+$ matrix element is written as
\begin{equation}
\label{DKpiMeissner}
\langle K^-\pi^+|\bar s\gamma^\mu(1-\gamma_5)c|D^+\rangle=\sum_{R=S,V}
\langle K^-\pi^+|R\rangle P_R\langle R|\bar s\gamma^\mu(1-\gamma_5)c|D^+\rangle\ ,
\end{equation}
where we assumed that only scalar and vector intermediate resonances propagate.
Tensor resonances are not included in the sum since the $K_2^\ast(1430)$
is seen to contribute less than 1\% \cite{Amsler:2008zzb}.
In Eq.~(\ref{DKpiMeissner}), $\langle K^-\pi^+|R\rangle$ is the coupling of $K\pi$ to the resonance and $P_R$ stands for the propagation of that resonance.
The same decomposition is possible for the matrix element which define the scalar and vector form factors.
Our aim is to substitute the products $\langle K^-\pi^+|R\rangle P_R$,
usually involving Breit-Wigner parametrisations,
by expressions based on the relevant form factors.

For the scalar case, let us take for instance the contribution of $K^\ast_0(1430)$ alone and write
\begin{equation}
\label{DKpiMeissnerK01430}
\langle K^-\pi^+|\bar s d|0\rangle=\frac{m_K^2-m_\pi^2}{m_s-m_d}F_0^{K\pi}(q^2)
=\langle K^-\pi^+|\bar K^\ast_0\rangle P_{\bar K^\ast_0}(q^2)
\langle\bar K^\ast_0|\bar s d|0\rangle\ ,
\end{equation}
where the matrix element $\langle \bar K^\ast_0|\bar s d|0\rangle$ defines the scalar decay constant.
Then,
\begin{equation}
\label{PiK0Kpi}
\Pi_{\bar K^\ast_0 K\pi}(q^2)
\equiv\langle K^-\pi^+|\bar K^\ast_0\rangle P_{\bar K^\ast_0}(q^2)
=\frac{1}{\langle\bar K^\ast_0|\bar s d|0\rangle}\frac{m_K^2-m_\pi^2}{m_s-m_d}F_0^{K\pi}(q^2)
\equiv\chi_{\bar K^\ast_0}F_0^{K\pi}(q^2)\ ,
\end{equation}
with $\chi_{\bar K^\ast_0}$ a pure number understood as a normalisation.
Hence, the contribution to the matrix element in Eq.~(\ref{DKpiMeissner}) is
\begin{equation}
\langle K^-\pi^+|\bar s\gamma^\mu(1-\gamma_5)c|D^+\rangle\Big|_{\bar K^\ast_0}
=\chi_{\bar K^\ast_0}F_0^{K\pi}(q^2)\langle\bar K^\ast_0|\bar s\gamma^\mu(1-\gamma_5)c|D^+\rangle\ .
\end{equation}
In order to make contact with Ref.~\cite{Diakonou:1989sf}
one can consider the function $\Pi_{\bar K^\ast_0 K\pi}$ in a Breit-Wigner parametrisation,
\begin{equation}
\label{PiK0KpiBW}
\Pi_{\bar K^\ast_0 K\pi}^{\rm BW}(q^2)=
\frac{g_{\bar K^\ast_0 K\pi}m_{\bar K^\ast_0}}
        {m_{\bar K^\ast_0}^2-q^2-i m_{\bar K^\ast_0}\Gamma_{\bar K^\ast_0}}\ ,
\end{equation}
recovering the scalar contribution in Eq.~(\ref{A1Diakonosv2}).
For the remaining matrix element we use
\begin{equation}
\label{DK01430FF}
{\setlength\arraycolsep{2pt}
\begin{array}{rcl}
\langle\bar K^\ast_0|\bar s\gamma^\mu(1-\gamma_5)c|D^+\rangle
&=&-i\left\{
\displaystyle
\left[(p_D+p_{\bar K^\ast_0})^\mu-\frac{m_D^2-m_{\bar K^\ast_0}^2}{q^2}q^\mu\right]
F_+^{D\bar K^\ast_0}(q^2)\right.\\[2ex]
&&
\displaystyle
\left.+\frac{m_D^2-m_{\bar K^\ast_0}^2}{q^2}q^\mu F_0^{D\bar K^\ast_0}(q^2)\right\}\ ,
\end{array}
}
\end{equation}
with $q=p_D-p_{\bar K^\ast_0}$.
Finally, we get the $K^\ast_0(1430)$ contribution to ${\cal A}_1$,
\begin{equation}
\label{A1Meissner1SR}
{\setlength\arraycolsep{2pt}
\begin{array}{l}
\langle K^-\pi^+_1|\bar s\gamma^\mu(1-\gamma_5)c|D^+\rangle\Big|_{\bar K^\ast_0}
\langle\pi^+_2|\bar u\gamma_\mu(1-\gamma_5)d|0\rangle\\[1ex]
\hspace{8em}
=f_\pi\chi_{\bar K^\ast_0}(m_D^2-m_{\bar K^\ast_0}^2)
F_0^{D\bar K^\ast_0}(m_\pi^2)F_0^{K\pi}(m_{K\pi_1}^2)\ .
\end{array}
}
\end{equation}
From Eqs.~(\ref{PiK0Kpi}) and (\ref{PiK0KpiBW}),
one can get an estimate of the absolute value of $\chi_{\bar K^\ast_0}$,
\begin{equation}
\label{chiK01430}
\chi_{\bar K^\ast_0}=\left|
\frac{\Pi_{\bar K^\ast_0 K\pi}^{\rm BW}(m_{\bar K^\ast_0}^2)}{F_0^{K\pi}(m_{\bar K^\ast_0}^2)}
\right|=
\frac{g_{\bar K^\ast_0 K\pi}}{\Gamma_{\bar K^\ast_0}(m_{\bar K^\ast_0}^2)}
\frac{1}{|F_0^{K\pi}(m_{\bar K^\ast_0}^2)|}=(4.4\pm 2.8)\ \mbox{GeV}^{-1}\ ,
\end{equation}
where the error includes only the uncertainty in $g_{\bar K^\ast_0 K\pi}$ and
$\Gamma_{\bar K^\ast_0}$.
For the numerical values we have used
$g_{\bar K^\ast_0 K\pi}=3.4\pm 1.9$,
obtained from ${\cal B}(K^\ast_0\to K\pi)=(93\pm 10)\%$,
$\Gamma_{\bar K^\ast_0}=270\pm 80$ MeV \cite{Amsler:2008zzb}, and
$|F_0^{K\pi}(m_{\bar K^\ast_0}^2)|=2.89$ from Ref.~\cite{Jamin:2006tj}. 

If more than one scalar resonance is exchanged then
\begin{equation}
\label{A1MeissnerNSR}
{\setlength\arraycolsep{2pt}
\begin{array}{l}
\displaystyle
\langle K^-\pi^+_1|\bar s\gamma^\mu(1-\gamma_5)c|D^+\rangle\Big|_S
\langle\pi^+_2|\bar u\gamma_\mu(1-\gamma_5)d|0\rangle\\[1ex]
\hspace{8em}
\displaystyle
=f_\pi\sum_S\left[\chi_S(m_D^2-m_S^2)F_0^{D S}(m_\pi^2)\right]F_0^{K\pi}(m_{K\pi_1}^2)\ .
\end{array}
}
\end{equation}
In Eqs.~(\ref{A1Meissner1SR}) and (\ref{A1MeissnerNSR}),
the scalar resonances are taken on-shell since it is assumed we are in the vicinity of these resonances and hence only small energy regions around the resonance poles are considered.
However, we want to describe the whole $K\pi$ invariant mass range.
For such a description, we propose the following ansatz for the scalar contribution to ${\cal A}_1$,
\begin{equation}
\label{A1MeissnerNSRq2}
{\setlength\arraycolsep{2pt}
\begin{array}{rcl}
\displaystyle
{\cal A}_1^S
&=&
\langle K^-\pi^+_1|\bar s\gamma^\mu(1-\gamma_5)c|D^+\rangle\Big|_S
\langle\pi^+_2|\bar u\gamma_\mu(1-\gamma_5)d|0\rangle\\[1ex]
&=&
\displaystyle
f_\pi\sum_S\left[\chi_SF_0^{D S}(m_\pi^2)\right](m_D^2-m_{K\pi_1}^2)F_0^{K\pi}(m_{K\pi_1}^2)
\\[1ex]
&\equiv&
f_\pi\chi_S^{\rm eff}(m_D^2-m_{K\pi_1}^2)F_0^{K\pi}(m_{K\pi_1}^2)\ ,
\end{array}
}
\end{equation}
where $\chi_S^{\rm eff}$ is a new normalisation constant that contains all the form factors and normalisations for the scalar resonances.
An estimate for $\chi_S^{\rm eff}$ is given by
\begin{equation}
\label{chiSeff}
\chi_S^{\rm eff}\geq\chi_{\bar K^\ast_0}F_0^{D\bar K^\ast_0}(m_\pi^2)=
(5.5\pm 3.5)\ \mbox{GeV}^{-1}\ ,
\end{equation}
where the value $F_0^{D\bar K^\ast_0}(m_\pi^2)=1.24\pm 0.07$ is taken from
Ref.~\cite{Cheng:2002ai}.
This value, obtained assuming that the form factor is saturated by the $D_s^+$ pole,
is consistent with $1.20\pm 0.07$ extracted directly from
$D^+\to\bar K^{\ast 0}_0\pi^+$ \cite{Cheng:2002ai}.
Since the estimate in Eq.~(\ref{chiSeff}) is a lower bound,  we prefer to leave $\chi_S^{\rm eff}$
as a free parameter of our analysis to be determined from the reported value of
${\cal B}(D^+\to K^-\pi^+\pi^+)$~\cite{Amsler:2008zzb}.

For the vector case, let us discuss in some detail the contribution of $K^\ast(892)$.
On one side, one takes the vector current matrix element in Eq.~(\ref{Kpiff}) and writes
\begin{equation}
\label{KpiK892}
{\setlength\arraycolsep{2pt}
\begin{array}{rcl}
\langle K^-\pi^+|\bar s\gamma^\mu d|0\rangle\Big|_{\bar K^\ast}
&=&
\displaystyle
\sum_{\rm pol.}
\langle K^-\pi^+|\bar K^\ast\rangle
P_{\bar K^\ast}(q^2)
\langle\bar K^\ast|\bar s\gamma^\mu d|0\rangle\\
&=&
\displaystyle
g_{\bar K^\ast K\pi}m_{\bar K^\ast}f_{\bar K^\ast}P_{\bar K^\ast}(q^2)
[(p_K-p_\pi)^\mu+\cdots]\ ,
\end{array}
}
\end{equation}
where $\langle K^-\pi^+|\bar K^\ast\rangle=g_{\bar K^\ast K\pi}\,\epsilon(q)\cdot (p_K-p_\pi)$,
$\langle\bar K^\ast|\bar s\gamma^\mu d|0\rangle=
-m_{\bar K^\ast}f_{\bar K^\ast}\epsilon^{\mu\ast}(q)$, with $q=p_K+p_\pi$, and
$\sum_{\rm pol.}\epsilon_\mu(q)\epsilon^\ast_\nu(q)=-g_{\mu\nu}+q_\mu q_\nu/m_{\bar K^\ast}^2$.
We have made explicit only the contribution of the vector transverse degrees of freedom.
The dots stand for the longitudinal degrees of freedom which can be shown to contribute to both the scalar and vector form factors.
However, for the sake of comparison, it is enough to consider the transverse part.
Comparing Eqs.~(\ref{Kpiff}) and (\ref{KpiK892}), one finds the equality
\begin{equation}
\Pi_{\bar K^\ast K\pi}(q^2)\equiv
g_{\bar K^\ast K\pi}m_{\bar K^\ast}P_{\bar K^\ast}(q^2)=
\frac{F_+^{K\pi}(q^2)}{f_{\bar K^\ast}}\equiv\chi_{\bar K^\ast}F_+^{K\pi}(q^2)\ .
\end{equation}
The former equality must be understood as a replacement of the $K^\ast (892)$ contribution by the vector form factor.
This replacement should be valid at least in the region around the resonance.
A direct estimate of $\chi_{\bar K^\ast}=(4.9\pm 0.2)\ \mbox{GeV}^{-1}$ is obtained using
$f_{\bar K^\ast}=(205\pm 6)$ MeV from 
$B(\tau^-\to K^\ast(892)^-\nu_\tau)=(1.20\pm 0.07)\%$ \cite{Amsler:2008zzb}.
On the other side, one has
\begin{equation}
\label{qDK892}
q_\mu\langle\bar K^\ast|\bar s\gamma^\mu(1-\gamma_5)c|D^+\rangle=
i(\epsilon^\ast\cdot q) 2 m_{\bar K^\ast}F_+^{D\bar K^\ast}(q^2)\ ,
\end{equation}
where the matrix element $\langle\bar K^\ast|\bar s\gamma^\mu(1-\gamma_5)c|D^+\rangle$
is written in general in terms of four different form factors \cite{Wirbel:1985ji}.
However, after contraction with $q=p_D-p_{\bar K^\ast}$ only the scalar form factor
$F_+^{D\bar K^\ast}$ remains\footnote{
In the notation of Ref.~\cite{Wirbel:1985ji},
$F_+^{D\bar K^\ast}$ corresponds to $A_0^{D\bar K^\ast}$.}.
Finally, the $K^\ast(892)$ contribution to ${\cal A}_1$ is written as
\begin{equation}
\label{A1Meissner1VR}
{\setlength\arraycolsep{2pt}
\begin{array}{l}
\langle K^-\pi^+_1|\bar s\gamma^\mu(1-\gamma_5)c|D^+\rangle\Big|_{\bar K^\ast}
\langle\pi^+_2|\bar u\gamma_\mu(1-\gamma_5)d|0\rangle\\[1ex]
\hspace{1em}
\displaystyle
=\sum_{\rm pol.}\langle K^-\pi^+_1|\bar K^\ast\rangle P_{\bar K^\ast}(m_{K\pi_1}^2)
\langle\bar K^\ast|\bar s\gamma^\mu(1-\gamma_5)c|D^+\rangle
\langle\pi^+_2|\bar u\gamma_\mu(1-\gamma_5)d|0\rangle\\[1ex]
\hspace{1em}
=f_\pi\Pi_{\bar K^\ast K\pi}(m_{K\pi_1}^2)N(m_{\bar K^\ast}^2)F_+^{D\bar K^\ast}(m_\pi^2)
=f_\pi\chi_{\bar K^\ast}N(m_{\bar K^\ast}^2)F_+^{D\bar K^\ast}(m_\pi^2) F_+^{K\pi}(m_{K\pi_1}^2)
\ .\\[1ex]
\end{array}
}
\end{equation}
In the Breit-Wigner parametrisation, the function $\Pi_{\bar K^\ast K\pi}$ corresponds to
\begin{equation}
\label{PiKKpi}
\Pi_{\bar K^\ast K\pi}^{\rm BW}(m_{K\pi_1}^2)=
\frac{m_{\bar K^\ast}g_{\bar K^\ast K\pi}}
        {m_{\bar K^\ast}^2-m_{K\pi_1}^2-i m_{\bar K^\ast}\Gamma_{\bar K^\ast}}\ ,
\end{equation}
again recovering the vector contribution in Eq.~(\ref{A1Diakonosv2}).

Considering the exchange of more than one vector resonance  Eq.~(\ref{A1Meissner1VR}) turns into
\begin{equation}
\label{A1MeissnerNVR}
{\setlength\arraycolsep{2pt}
\begin{array}{l}
\displaystyle
\langle K^-\pi^+_1|\bar s\gamma^\mu(1-\gamma_5)c|D^+\rangle\Big|_V
\langle\pi^+_2|\bar u\gamma_\mu(1-\gamma_5)d|0\rangle\\[1ex]
\hspace{8em}
\displaystyle
=f_\pi\sum_V\left[\chi_V N(m_V^2)F_+^{DV}(m_\pi^2)\right]F_+^{K\pi}(m_{K\pi_1}^2)\ .
\end{array}
}
\end{equation}
Analogously to the scalar case, we propose to take for the vector contribution to ${\cal A}_1$,
\begin{equation}
\label{A1MeissnerNVRq2}
{\setlength\arraycolsep{2pt}
\begin{array}{rcl}
\displaystyle
{\cal A}_1^V
&=&
\langle K^-\pi^+_1|\bar s\gamma^\mu(1-\gamma_5)c|D^+\rangle\Big|_V
\langle\pi^+_2|\bar u\gamma_\mu(1-\gamma_5)d|0\rangle\\[1ex]
&=&
f_\pi
\displaystyle
\sum_V\left[\chi_V F_+^{DV}(m_\pi^2)\right]N(m_{K\pi_1}^2)F_+^{K\pi}(m_{K\pi_1}^2)\\[1ex]
&\equiv &
f_\pi\chi_V^{\rm eff}N(m_{K\pi_1}^2)F_+^{K\pi}(m_{K\pi_1}^2)\ .
\end{array}
}
\end{equation}
A lower bound for $\chi_V^{\rm eff}$ is obtained as
\begin{equation}
\label{chiVeff}
\chi_V^{\rm eff}\geq\chi_{\bar K^\ast}F_+^{D\bar K^\ast}(m_\pi^2)=
(4.6\pm 0.9)\ \mbox{GeV}^{-1}\ ,
\end{equation}
where the error takes into account the different results for
$F_+^{D\bar K^\ast}(m_\pi^2)\simeq F_+^{D\bar K^\ast}(0)$ extracted
from recent analyses.  The value $F_+^{D\bar K^\ast}(0)=0.76$ is found
in a quark model calculation \cite{Melikhov:2000yu} and a lattice
simulation \cite{Abada:2002ie}.  This value contrasts with $F_+^{D\bar
  K^\ast}(0)=1.12$ found in Ref.~\cite{Fajfer:2005ug} using limits of
large energy effective theory and heavy quark effective theory.  In
any case, we like better to leave $\chi_V^{\rm eff}$ as a second free
parameter to be fixed from the experimental value of ${\cal
  B}(D^+\to\bar K^\ast_0(892)\pi^+)+{\cal B}(D^+\to\bar
K^\ast_0(1680)\pi^+)$ \cite{Amsler:2008zzb}.

In Section.~\ref{results}, we perform a rather exhaustive numerical
analysis of our model and the models of
Refs.~\cite{Bediaga:1996ue,Diakonou:1989sf}.  For the sake of clarity,
our model is defined by the amplitude in Eq.~(\ref{amplitude})
resulting from the sum of colour-allowed scalar and vector
contributions, Eqs.~(\ref{A1MeissnerNSRq2}) and
(\ref{A1MeissnerNVRq2}) respectively,
\begin{equation}
\label{A1final}
{\cal A}_1=f_\pi\chi_S^{\rm eff}(m_D^2-m_{K\pi_1}^2)F_0^{K\pi}(m_{K\pi_1}^2)
+f_\pi\chi_V^{\rm eff}N(m_{K\pi_1}^2)F_+^{K\pi}(m_{K\pi_1}^2)\ ,
\end{equation}
and the colour suppressed contribution ${\cal A}_2$ in Eq.~(\ref{A2}).
It is worth mentioning, however, that the total scalar amplitude of
our model must be re-phased by some amount in order to carry out a fair comparison with experimental
results, see Eq.~(\ref{Shift}) for
details.  We denote this model as our final model.

%%%% NUMERICAL RESULTS   %%%%%%%%%%%%%%%%%%%%%%%%%%%%%%%%%%%%%%%%%%%%%%
\section{Numerical Results}
\label{results}

In this section we shall collect all the numerical results arising
from the models discussed in the previous section and compare them
with the  experimental results available.  Concerning the branching
ratios, we shall take the PDG averages~\cite{Amsler:2008zzb} shown as
the second column of Table~\ref{tab1}. From this table we learn that
(i) the contribution of the $K^-\pi^+\pi^+$ mode to $D^+$ decays is
important and accounts for about 10\% of these decays, (ii) the decay
is strongly dominated by $(K^-\pi^+)$ pairs in the $S$ wave, (iii)
although less important the vector $K^*(892)$ also gives a sizable
contribution, and (iv) the branching ratios of submodes containing the
next vector and the tensor resonances are fairly small.

\begin{table}[!ht]
\begin{ruledtabular}
\caption{ World average for the relevant branching ratios as reported
  by the Particle Data Group~\cite{Amsler:2008zzb} and results for the three
  models discussed in Section \ref{amplitudes}.  $(K^-\pi^+)_{S,P}$ denote $K^-\pi^+$ pairs
in   $S$ or $P$ wave. }\label{tab1}
\begin{tabular} {l c c c c }
Mode  & World Average~\cite{Amsler:2008zzb} & Model from Ref.~\cite{Diakonou:1989sf}&$\A_2$ only   &  Our Model \\ 
\hline
$D^+ \to K^-\pi^+\pi^+$  &  (9.22 $\pm$ 0.21) \% &  0.63\% &  $3.17\,$ \%      & Fixed\\
$D^+ \to (K^-\pi^+)_{S}\,\pi^+$  & (7.54 $\pm$ 0.26) \% &   $\cdots$& $0.27$ \%    & $(7.6 \pm 0.2)$ \%   \\
$D^+ \to (K^-\pi^+)_{P}\,\pi^+$  & $\cdots$ &       $\cdots$& $2.84$  \%          & Fixed \\
$D^+ \to K_0^{*}(1430)  \,\pi^+$& $\cdots$        &  $0.016\%$             &  $\cdots$         & $\cdots$     \\
$D^+ \to K^*(892) \,\pi^+$    &     (1.22 $\pm$ 0.09) \% &  $0.5\%$        &$\cdots$ & $\cdots$  \\
$D^+ \to K^{*}(1680) \,\pi^+$   &     (0.16 $\pm$ 0.06) \% & $\cdots$&    $\cdots$        &$\cdots$\\
$D^+ \to K_2^{*}(1430) \,\pi^+$   &     (0.030 $\pm$ 0.008) \%& $\cdots$&    $\cdots$      &$\cdots$\\
\end{tabular}
\end{ruledtabular}
\end{table}

Often, the branching ratios for the submodes are estimated from the
experimental fit to the Dalitz plot through fit fractions. These
fractions quantify the weight of the $i$-th  component of the amplitude to the
final result as
\beq
f_i = \frac{\int_{\mathcal{D}} \, d m_{K\pi_{1}}^2 d  m_{K\pi_{2}}^2\, |\mathcal{A}_i|^2}{\int_{\mathcal{D}} \,  d m_{K\pi_{1}}^2 d m_{K\pi_{2}}^2 \, |\sum_j \mathcal{A}_j|^2}\ .
\eeq
In this formula $i$ represents a submode that can be a resonance or
the sum of an entire partial wave and $\mathcal{D}$ denotes that the
integrals are to be evaluated over the whole Dalitz plot
(see Ref.~\cite{Amsler:2008zzb}). The fit fractions from the analyses
of~Refs.~\cite{Aitala:2002kr,Aitala:2005yh,Pennington:2007se,Focus2009,Bonvicini:2008jw}
are shown in Table~\ref{tab2} along with the results of our model,
discussed in the remainder of this section. Experimental groups have
used different models to fit the Dalitz plot.  In
Ref.~\cite{Aitala:2002kr} the isobar model was employed and the
contribution from the $\kappa$ was included as a Breit-Wigner
function. In Ref.~\cite{Pennington:2007se} a $K$-matrix model was used
for the $S$ wave. The results from
Refs.~\cite{Aitala:2005yh,Bonvicini:2008jw,Focus2009} are obtained
using a quasi-model-independent bin-by-bin analysis for the $S$ wave
introduced in Ref.~\cite{Aitala:2005yh}. In Ref.~\cite{Bonvicini:2008jw}, a $(\pi^+\pi^+)_{I=2}$ amplitude is also included in the model and is found to give a sizable contribution.

Finally, a comprehensive account of the decay should be able to
reproduce not only the known branching ratios and fit fractions but
also the detailed shape of the Dalitz plot. This is discussed for our
final model at the end of this section.

\begin{table*}[!ht]
\begin{ruledtabular}
%\begin{center}
\caption{ Fit fractions (in $\%$) for the different submodes of the
  decay $\DKpp$.  From Ref.~\cite{Bonvicini:2008jw} we quote the
  values for the quasi-model-independent analysis given in their
  Table VII. Results marked with an asterisk are 
  the sum of all contributions to a given partial wave. They   do not take into account interference effects and were not quoted in the original works. The errors in the results of our model take into account the uncertainties in $a_1$ and $a_2$, Eq.~(\ref{a12}).  }\label{tab2}
\begin{tabular} {l c c c c c c }
& { E791 ('02)~\cite{Aitala:2002kr}}   & E791 ('06)~\cite{Aitala:2005yh}& FOCUS ('07)~\cite{Pennington:2007se}&FOCUS ('09)~\cite{Focus2009} & CLEO~\cite{Bonvicini:2008jw}   & { Our Model }     \\ 
\hline
NR & $13\pm 7 $ & $\cdots$ & $\cdots$ &  $\cdots$  &$\cdots$   &$\cdots$  \\
$\kappa$& $48\pm 13$  &$\cdots$ & $\cdots$&  $\cdots$   &$\cdots$  & $\cdots$  \\
$K_0^*(1430)$ & $ 12.5 \pm 1.5$&$\cdots$  & $\cdots$ &  $\cdots$   & $13.3\pm 0.6$  & $\cdots$ \\
$ K^*(892)$  & $12.3\pm 1.3 $ & $11.9\pm 2.0$  & $13.6 \pm 1.0$  & $12.4\pm 0.5$    &$9.8\pm 0.5$  &$\cdots$  \\
$K^{*}(1410)$ & $\cdots$& $\cdots$ &$0.48 \pm 0.27$  &  $\cdots$   &$\cdots$ &$\cdots$  \\
$K^{*}(1680)$ & $ 2.5 \pm 0.8$& $1.2\pm 1.3$ &$1.9 \pm 0.8$  &  $1.8\pm 0.8$    &$0.20 \pm 0.12$  & $\cdots$  \\
$K_2^{*}(1430)$ &$0.5\pm 0.2$ & $0.2\pm 0.1 $ &$ 0.39 \pm 0.10$& $0.58\pm 0.12$    &$0.20\pm 0.04$ &$\cdots$  \\
$(K^-\pi^+)_S$ &$(73 \pm  15  )^*$ &$78.6 \pm 2.3$ & $83.2\pm 1.5$&   $80.2\pm 1.4$     &$83.8\pm 3.8$  & $82.0\pm 0.3$  \\
$(K^-\pi^+)_P$ &$(14.8 \pm 1.5)^* $ &$(13.1\pm 2.4)^*$  &$(16.0\pm 1.3)^*$ & $(14.2\pm 0.9)^*$      & $(10.0\pm 0.5)^*$ & $15.0\pm 0.2$  \\
$(\pi^+ \pi^+)_{I=2}$  & $\cdots$ & $\cdots$ & $\cdots$ &  $\cdots$  &$15.5\pm 2.8$   &$\cdots$    \\
$\sum_i f_i$ & 88.6 &91.9 &99.57 &   94.93    & $122.8$   &97.0 \\
\end{tabular}
%\end{center}
\end{ruledtabular}
\end{table*}

\subsection{Previous models in the literature}

Here,  we update the results of two models found in the
literature for the decay under study~\cite{Diakonou:1989sf,Bediaga:1996ue}. In both cases the
description of the weak decay is based on the effective weak
Hamiltonian and therefore it is simple to make contact with our model.
%The model by D and D ::
We begin considering the model presented in
Ref.~\cite{Diakonou:1989sf} where $\mathcal{A}_1$ is described by
Eq.~(\ref{A1Diakonosv2}) and $\mathcal{A}_2$ is given by
Eq.~(\ref{A2}).  We have updated the values of the relevant constants
for the form factors and for the Breit-Wigner parameters as compared
with the original work and calculated branching ratios and fit
fractions from this model.  The outcome of this exercise is shown in
Tables~\ref{tab1} and~\ref{tab3}. The total branching ratio obtained
is about a factor of 15 smaller than the world average. Moreover, in
Table~\ref{tab3} we show that this result is largely dominated by the
$K^*(892)$ with a fraction of 86.1\% of the total result.  The
$S$-wave component is represented by the $K_0^*(1430)$ alone and
accounts for $10.7\%$ of the result.  Therefore, the model fails to
reproduce the absolute branching fractions of Table~\ref{tab1} and the
strong dominance of the $S$ wave that is evident from
Tables~\ref{tab1} and~\ref{tab2}. As a last comment, note that we
employed the central values for $a_1$ and $a_2$ given in
Eq.~(\ref{a12}). Shifts within uncertainties in these values could
produce sizable changes in the branching fractions. However, since the
general picture of this model does not agree with the known $S$-wave
dominance, we do not attempt to fine-tune these values in the case at
hand.

%The model with A_2 only ::
Before turning to our final model it is worth investigating the
suggestion of Ref.~\cite{Bediaga:1996ue}. In this work, the authors
advocated that $\A_2$ should give the dominant contribution since the
colour-allowed topology appears multiplied by a factor of $ f_\pi
m_\pi^2$. Following this suggestion, we ignore for the moment the
colour-allowed topology. In the amplitude $\A_2$ the $K\pi$ form
factors enter manifestly and it is straightforward to introduce the
ones from Refs.~\cite{Jamin:2006tj,Boito:2008fq}, as shown in
Eq.~(\ref{A2}). The use of these form factors improves the description
of FSIs as compared with Ref.~\cite{Diakonou:1989sf} incorporating
constraints from  analyticity and unitarity.
 The numerical results from this
model are shown Tables~\ref{tab1} and~\ref{tab3}. 
The total branching ratio is now about a factor of 3 smaller than the
experimental average. Nevertheless, the dominant contribution is again
given by the $P$ wave that accounts for $89.6\%$ of the total
result. Since the result for $\A_2$ is unambiguous and we employed
state-of-the-art form factors we are led to the conclusion that $\A_1$
must be taken into account. As a matter of fact, it is now transparent
that the large $S$-wave contribution originates precisely in the
colour-allowed topology.

\begin{table}[!ht]
\begin{ruledtabular}
\caption{Results for the fit fractions  (in \%) arising from the three models 
described in the text.  }\label{tab3}
\begin{tabular} {l c c c }
Mode  &  Model from Ref.~\cite{Diakonou:1989sf}  & $\A_2$ only &Our Model\\ 
\hline
NR & $\cdots $ & $\cdots $ &$\cdots $ \\
$K_0^*(1430)$ & $ 10.7 $&  $\cdots $ &$\cdots $\\
$ K^*(892)$  & $86.1 $ & $\cdots $ &$\cdots $\\
$(K^-\pi^+)_S$ &$\cdots$ & $8.5 $ & $82.0\pm 0.3$  \\
$(K^-\pi^+)_P$ &$\cdots$  & $89.6$& $15.0\pm 0.2$ \\
$\sum_i f_i$ & 96.8 &  98.1   & 97.0 \\
\end{tabular}
\end{ruledtabular}
\end{table}

%% RESULTS FROM THE FINAL MODEL

\subsection{Our model}

Let us now investigate in detail the numerical results for our final
model which includes the contribution of both $\A_1$ and $\A_2$
topologies. The corresponding expressions are given in Eqs.~(\ref{A2})
and~(\ref{A1final}). We begin by considering the $S$-wave description
which is, in our opinion, the main aspect of the problem. On the
experimental side, in 2006, E791 introduced a new type of Dalitz plot
analysis~\cite{Aitala:2005yh} where, instead of modelling the $K\pi$
$S$ wave, its absolute value and phase are determined in a bin-by-bin
basis directly from data. This is done assuming a reference amplitude,
customarily that of the $K^*(892)$.  The analysis was repeated by
CLEO~\cite{Bonvicini:2008jw} and FOCUS~\cite{Focus2009} with similar
results. It is important to remark that this framework can only be
considered as $quasi$-model-independent (QMI) since the $P$ and $D$
waves are still described by their isobar expressions. Nevertheless,
since the isobar prescription seems to be more accurate for these
latter waves, one can expect that the results have little model
dependence. Therefore, the results of the QMI analyses of
Refs.~\cite{Aitala:2005yh,Focus2009,Bonvicini:2008jw} are the better
source of empirical information about the $K\pi$ $S$-wave amplitude in
$\DKpp$.

\begin{figure}[!ht]
\includegraphics[width=0.75\columnwidth,angle=0]{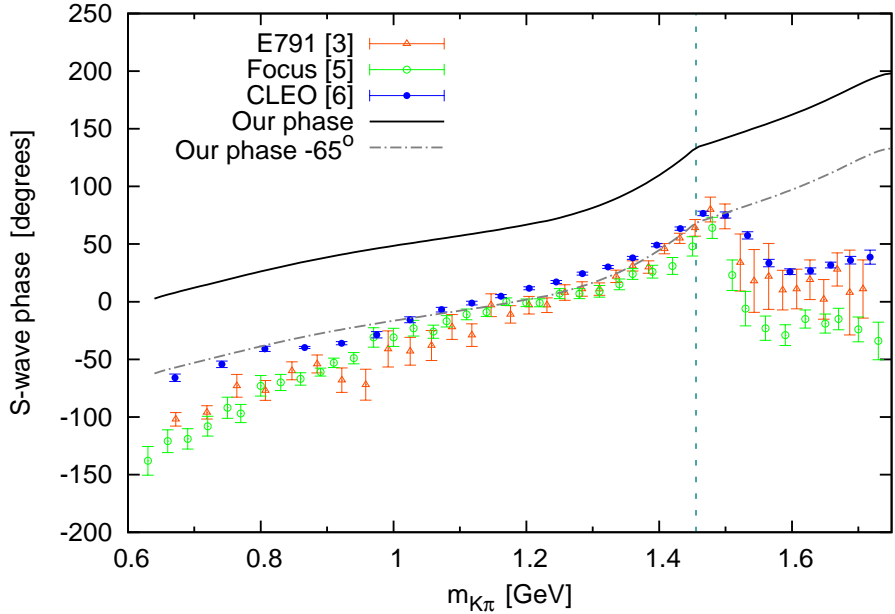}
\caption{{(colour online). $S$-wave phases from the QMI analyses of Refs.~\cite{Aitala:2005yh,Bonvicini:2008jw,Focus2009}. The solid line is the phase of our $S$-wave amplitude with $\alpha=0^\circ$ in Eq.~(\ref{Shift}), whereas the dot-dashed line is  the $S$-wave phase  with $\alpha=-65^\circ$. The dashed line delimits the $K\eta'$ threshold. } } \label{Phase}
\end{figure}

  The QMI measurement of the $S$-wave phase can be used to test
  whether Watson's theorem~\cite{Watson} holds for the three-body
  decay in question. The theorem states that, in the elastic domain,
  the $K\pi$ $S$ wave would exhibit the corresponding $K\pi$
  scattering phase shift. However, this is valid only in the absence
  of genuine three-body effects. Therefore, in $\DKpp$, the empirical
  $S$-wave phase could be distorted as compared with the scattering
  one due to interactions of the resonant $K\pi$ pair with the
  bachelor pion. In our model, the $S$-wave FSIs are described by the
  $K\pi$ scalar form factor of Ref.~\cite{Jamin:2006tj} in a quasi
  two-body approach, {\it i.e.}, we assume that the $K\pi$ pairs in
  Eq.~(\ref{amplitude}) form an isolated system and do not interact with the
  bachelor pion. Moreover, the form factor of
  Ref.~\cite{Jamin:2006tj} is obtained from dispersion relations that
  fix its phase to be the scattering one within the elastic
  region~\cite{Jamin:2001zq}. Consequently, our $S$-wave amplitude has
  the $K\pi$ $I=1/2$ scattering phase up to roughly 1.45 GeV where the
  $K\eta'$ channel starts playing a role. We compare in
  Fig.~\ref{Phase} the experimental results from
  Refs.~\cite{Aitala:2005yh,Bonvicini:2008jw,Focus2009} with the phase
  of our $S$ wave. The high-statistics results of CLEO collaboration
  have the smallest errors. One observes from Fig.~\ref{Phase} that
  the QMI phases start at negative values ranging from
  -60$^\circ$~\cite{Bonvicini:2008jw} to -145$^\circ$~\cite{Focus2009}
  whereas our phase evolves from $0^\circ$ up to about $200^\circ$
  (modulo $\pi$) in the allowed phase space. Since we are dealing with
  a production experiment, a global phase difference is expected as
  compared with scattering results~\cite{Pennington:2007se}.
  Therefore, we allow for a global phase shift $\alpha$ in our
  $S$-wave amplitude\footnote{The results of the model are sensitive
    only to the phase difference between the $S$ and $P$
    waves. Therefore, $\alpha$ can be considered as a global phase
    difference between the two waves.}:
\beq
\label{Shift}
\A_S(m_{K\pi_{1}}^2,m_{K\pi_{2}}^2)\rightarrow e^{i\alpha}\, \A_S(m_{K\pi_{1}}^2,m_{K\pi_{2}}^2)\ .
\eeq
In Fig.~\ref{Phase}, we also
plot as the dot-dashed line the phase of our amplitude  shifted by
$\alpha=-65^\circ$. With this shift, we see that up to 1.5 GeV CLEO's
results and ours share a remarkably similar dependence on energy\footnote{The
$S$-wave phase of the form factor of Ref.~\cite{Jamin:2006tj} exhibits around 1.8~GeV a deep similar to the
one observed in the experimental results of Fig.~\ref{Phase}.}.  The
results of E791 and FOCUS seem to have a somewhat different energy
dependence, although they have larger error bars due to smaller
statistics. Inspired by the inspection of Fig.~\ref{Phase}, we consider  as
our final model the one given by Eqs.~(\ref{A2})
and~(\ref{A1final}) with a shift of $\alpha=-65^\circ$ in the $S$-wave phase as defined 
in Eq.~(\ref{Shift}). We will discuss further  consequences of this shift below.

In order to compare the absolute value of our $S$ wave amplitude with
experimental data, we need fix the only two free parameters that occur
in our model, namely the normalisation constants $\chi_S^{\rm eff}$
and $\chi_V^{\rm eff}$. Estimates for the normalisations were given in
Eqs.~(\ref{chiSeff}) and~(\ref{chiVeff}) but in order to perform a
careful comparison with experimental results we choose to refine these
values.  With that aim, we employ the following strategy. The constant
$\chi_V^{\rm eff}$ is fixed in order to reproduce the value of the sum of
all vector submodes\footnote{This procedure does not take into account
  possible interference effects. However, these effects are likely to
  be small since the resonances are relatively narrow. Furthermore, to
  the best of our knowledge, there is no experimental value for the
  total $P$-wave branching ratio.} in the second column of
Table~\ref{tab1}. Then, we fix the scalar normalisation $\chi_S^{\rm
  eff}$ requiring the total branching ratio from our model to match
the world average of Table~\ref{tab1}. Taking the central values for
$a_1$ and $a_2$ given in Eq.~(\ref{a12}) this procedure gives
\beq
\chi_S ^{\rm eff}= 4.9 \pm 0.4 \,\,\,\mbox{GeV}^{-1} \ , \qquad
\chi_V^{\rm eff} = 4.4   \pm 0.6 \,\,\,\mbox{GeV}^{-1}\ ,
\eeq
in good agreement with our estimates in Eqs.~(\ref{chiSeff})
and~(\ref{chiVeff}). The uncertainties take into account the error in
$a_1$ and $a_2$ which dominate by far as compared to the relatively
small errors of the world averages of Table~\ref{tab1}.  We are now in
a position to compute the scalar branching ratio, shown in
Table~\ref{tab1}, as well as the fit fractions of the total vector and
scalar contributions which are shown in Tables~\ref{tab2}
and~\ref{tab3}. The model reproduces the dominant $S$-wave
contribution and gives fit fractions in fair agreement with the
experimental results.

\begin{figure}[!ht]
\includegraphics[width=0.75\columnwidth,angle=0]{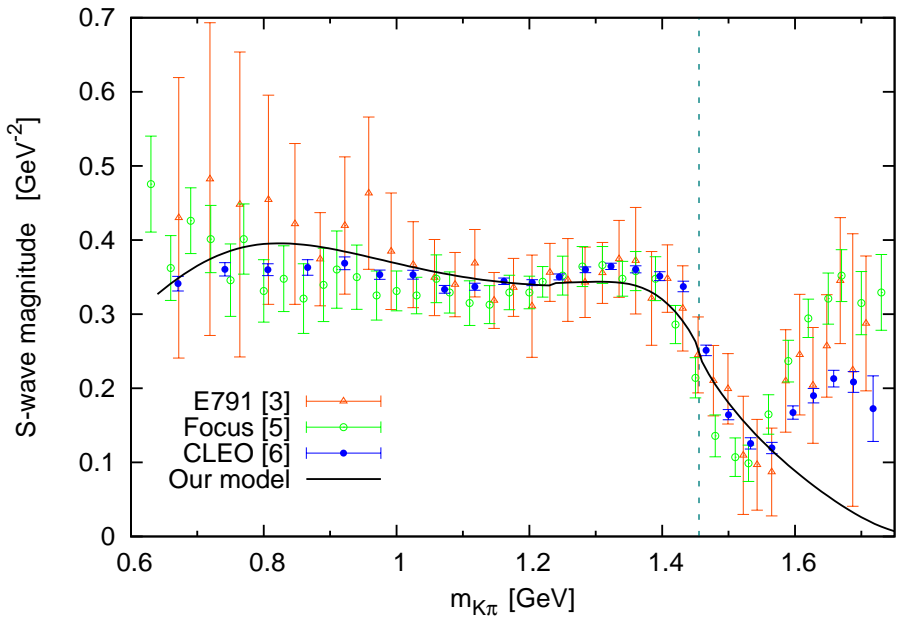}
\caption{{(colour online). Absolute value of the $S$ wave  measured in Refs.~\cite{Aitala:2005yh,Bonvicini:2008jw,Focus2009}  compared with our model.  The amplitudes are normalised according to Eq.~(\ref{SwaveNorm}). The dashed line delimits the $K\eta'$ threshold.} } \label{Swave}
\end{figure}

We can now compare the absolute value of our $S$-wave amplitude with experimental results
from the QMI analyses. However,
since in isobar-like analyses the fit is sensitive only to the
relative weights of the amplitudes, in order to compare the
measurements with our result we need perform a
normalisation. We define a normalised $S$-wave amplitude by
\beq
\A_S^{\mbox{\tiny Norm}}( m_{K\pi_{1}}^2, m_{K\pi_{2}}^2) = \frac{\A_S}{\left(\int_{\mathcal{D}} \, d  m_{K\pi_{1}}^2 d  m_{K\pi_{2}}^2\, | \mathcal{A}_S|^2\right)^{1/2} }\ .
\label{SwaveNorm}
\eeq
This amplitude, by construction, is  free of any global constants that
appear in  $\A_S$ and has dimension  of [Energy]$^{-2}$. Interpolating
the       results      from       the       tables      found       in
Refs.~\cite{Aitala:2005yh,Bonvicini:2008jw,Focus2009} we can calculate
the  normalised $S$  wave for  each experiment.  We repeated  the same
procedure for  our total $S$-wave  amplitude. The QMI results  for the
$S$  wave are  compared with  our  model in  Fig.~\ref{Swave}.  Up  to
$1.55$~GeV the agreement of  our results with the experimental ones is
remarkable.

\begin{figure}[!ht]
\begin{center}
\includegraphics[width=0.45\columnwidth,angle=0]{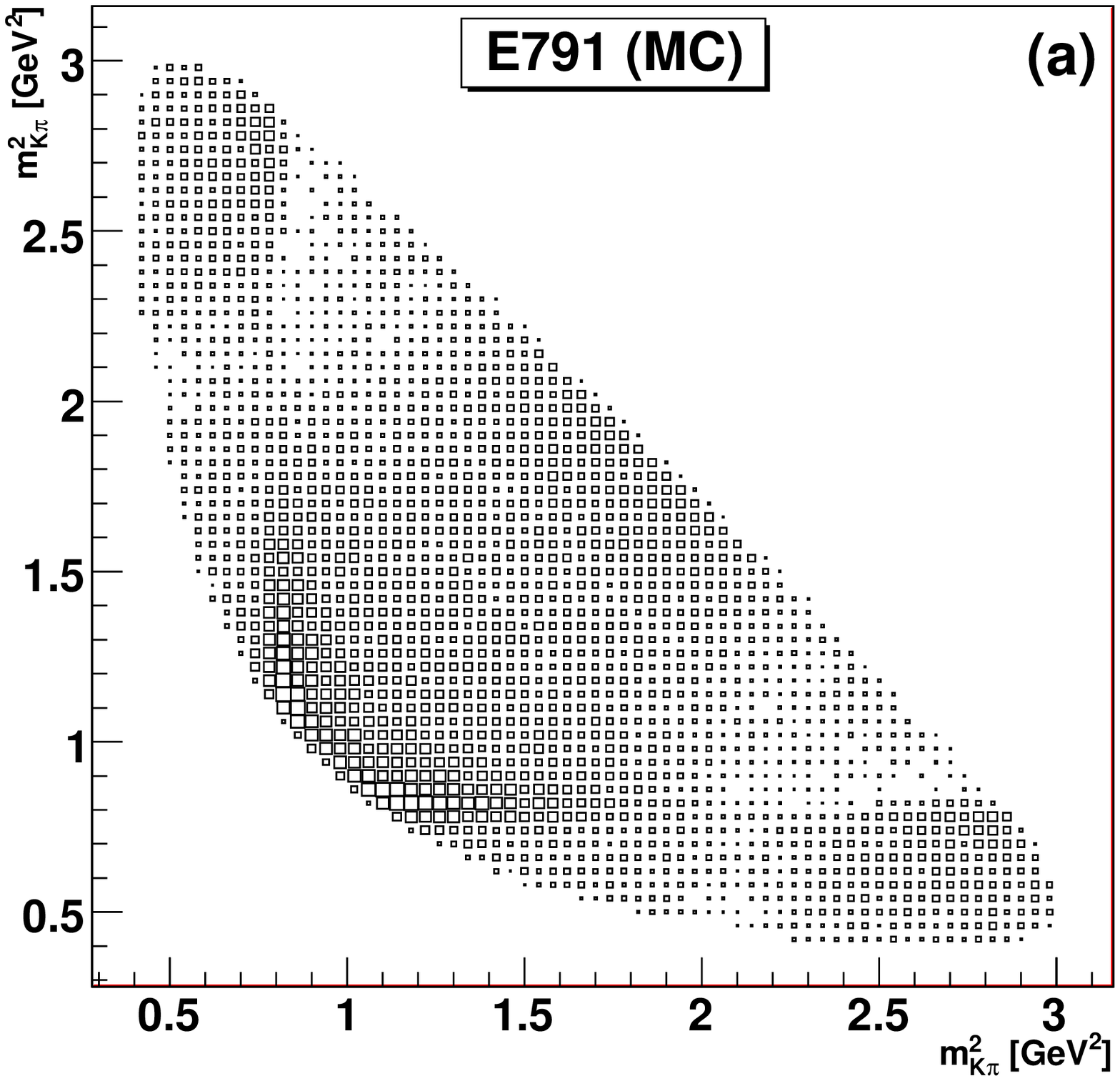}
\includegraphics[width=0.45\columnwidth,angle=0]{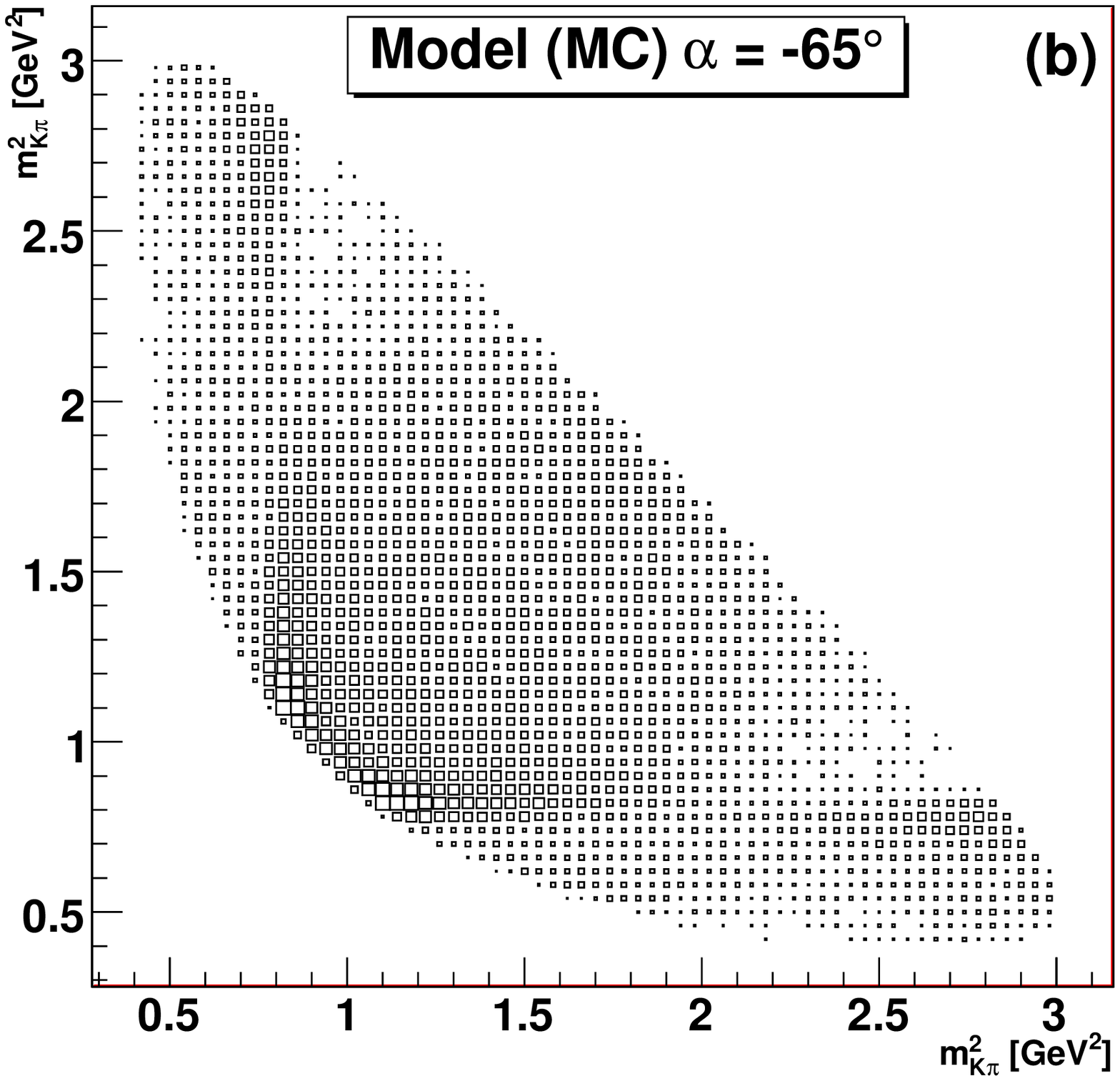}
\caption{{(a) Monte Carlo simulation for the Dalitz plot of the E791 original analysis~\cite{Aitala:2002kr} (b) Same for our model with a global shift of $-65^\circ$ degrees in the $S$-wave phase (see text and Fig.~\ref{Phase}).  The number of independent events is 14185, which correspond to the estimate of the signal events in Ref.~\cite{Aitala:2002kr}.  } }\label{Dalitz}
\end{center} 
\end{figure}

\begin{figure}[!ht]
\begin{center}
\includegraphics[width=0.55\columnwidth,angle=0]{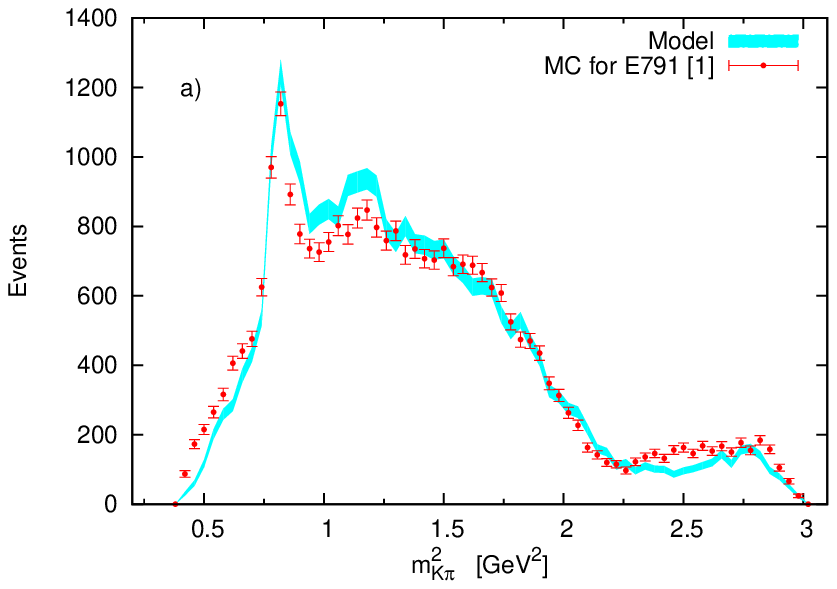}
\includegraphics[width=0.45\columnwidth,angle=0]{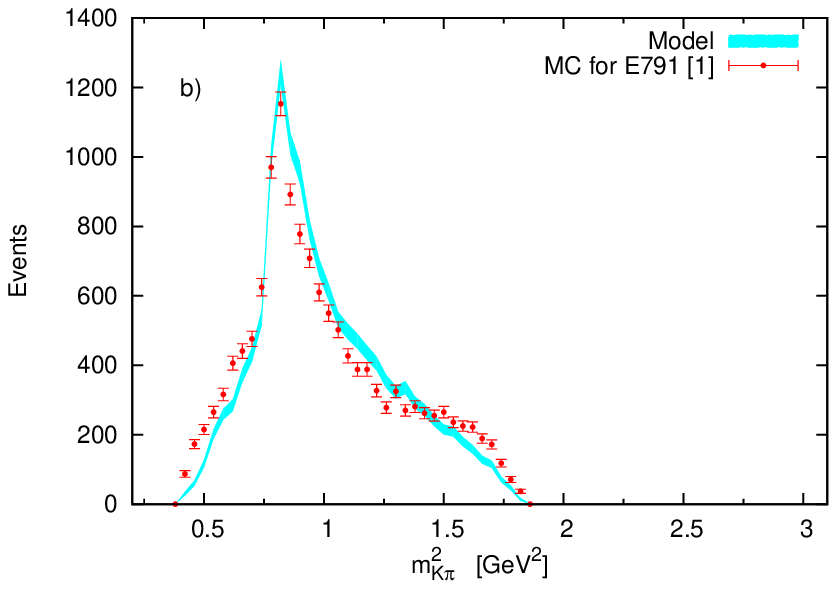}
\includegraphics[width=0.45\columnwidth,angle=0]{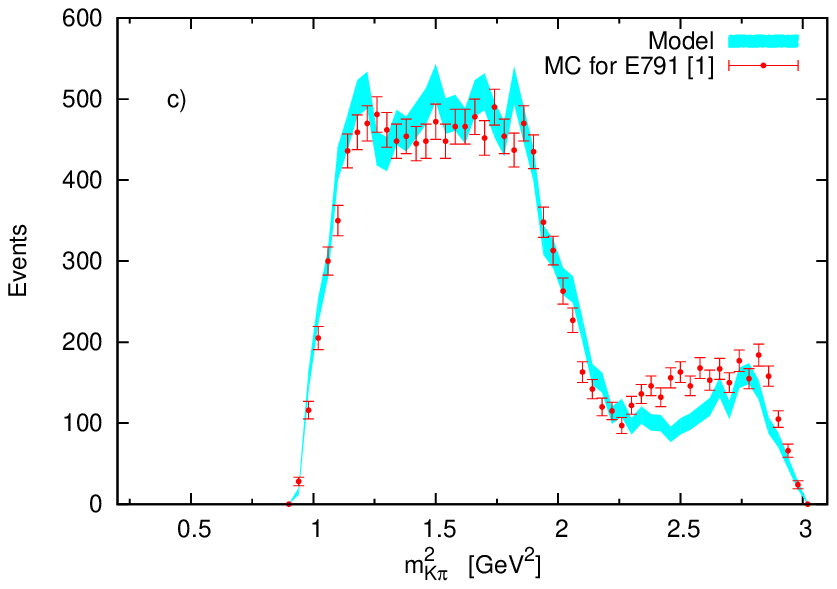}
\caption{{(colour online). Projections from the MC generated Dalitz plots of Figs.~\ref{Dalitz}a and~\ref{Dalitz}b. The error bars and the bands represent solely statistical fluctuations. (a) Total projection, (b) high-energy projection, (c) low-energy projection.} } \label{Projecs}\end{center} 
\end{figure}

 Finally, we can perform a Monte Carlo (MC) simulation to obtain a
 Dalitz plot from our model and compare the diagram and its
 projections with experimental results.  For the lack of a true data
 set, we resort to a MC simulation of the original E791
 data~\cite{Aitala:2002kr}. Reproducing their fit function, we
 generated a symmetrised Dalitz plot with 14185 independent signal
 events which corresponds to $6\%$ of background contamination in the
 total sample~\cite{Aitala:2002kr}. The obtained diagram is shown in
 Fig.~\ref{Dalitz}a.  Then we performed the same exercise for our
 model and the result is shown in Fig.~\ref{Dalitz}b.  It is important
 to remark that the shape of the Dalitz plot is related to the global
 phase shift of Eq.~(\ref{Shift}).  In the words of
 Ref.~\cite{Aitala:2005yh}, the asymmetry in the $K\pi$ $P$-wave bands
 reflects the value of $\alpha$.  We have checked that taking
 $\alpha=0^\circ$ in Eq.~(\ref{Shift}) reverses the observed
 asymmetry, {\it i.e.}, the high-energy part of the Dalitz is more populated
 than the low-energy corner.  Consequently, we confirm the finding of
 Ref.~\cite{Aitala:2005yh}: the asymmetry pattern in the Dalitz plot
 is a direct consequence of the global phase difference between the
 $S$- and $P$-wave phases. Finally, in Fig.~\ref{Projecs} we show
 the projections of the diagrams of Figs.~\ref{Dalitz}a
 and~\ref{Dalitz}b. The results for our model with $\alpha=-65^\circ$
 and the simulated E791 data agree quite well. The discrepancy in
 Fig.~\ref{Projecs}a around 1~GeV$^2$ is due to the interference
 pattern between the $S$ and $P$ waves. This could be fixed through a
 fit to real data, which would give a refined value for $\alpha$.  One
 also sees around 2.5~GeV$^2$ a second discrepancy, seen in both
 Figs.~\ref{Projecs}a and~\ref{Projecs}c, that is a consequence of the
 disagreement of our $S$ wave with respect to the experimental ones
 for $m_{K\pi}>1.45$ GeV, as shown in Fig.~\ref{Swave}. Small
isospin-breaking effects in the $P$-wave are to be expected as well,
since the vector $K\pi$ form factor employed here was obtained from
$\tau^-\to K\pi\, \nu_{\tau}$ decay data~\cite{Boito:2008fq} where the
 charged vector resonances intervene.

% Figures Dalitz plots

% Figures projections 

As a final comment, we remark that we do not include  $(\pi^+\pi^+)_{I=2}$ interactions in our model. 
Within the framework employed here this contribution does not appear. Since the inclusion of an ad-hoc $I=2$ amplitude would downgrade  the model, we prefer to consider only the $I=1/2$ FSIs.  Additionally,   $(\pi^+\pi^+)_{I=2}$ scattering is entirely non-resonant~\cite{Amsler:2008zzb} with a  slow variation of the corresponding phase shift~\cite{I2}, indicating that interactions in this channel are weak.
Furthermore, from an experimental point of view, the need for the $I=2$ amplitude is not well established and requires further confirmation   (see Table~\ref{tab2}).

%%%%%%%%%%%% SUMMARY AND DISCUSSION

\section{Summary and discussion}
\label{summary}

We have presented a model aimed at describing the decay $\DKpp$.  The
weak amplitude is described within the effective Hamiltonian framework with the
hypothesis of factorisation.  The $K\pi$ hadronic FSIs are treated in
a quasi two-body approach by means of the well defined scalar and
vector $K\pi$ form factors, thereby imposing  analyticity, unitarity
and chiral symmetry constraints. We used the experimental values for
the total and $P$-wave branching ratios to fix the two free parameters
in the model. The relative global phase difference between the $S$ and $P$ waves
was fixed phenomenologically using the experimental results of
Ref.~\cite{Bonvicini:2008jw}.

The use of the $K\pi$ scalar form factor is shown to provide a good
description of the $S$-wave FSIs.  Both the modulus and the phase of
our $S$ wave compare well with experimental data up to
$m_{K\pi}\lesssim 1.5$~GeV.  It is worth mentioning that the form
factor we used has a pole that can be identified with the
$\kappa$. Furthermore, the model is able to reproduce the experimental
fit fractions and the total $S$-wave branching ratio. Finally, the
Dalitz plot arising from the model agrees with a MC simulated
data set.

The main hypotheses of our model are the factorisation of the weak
decay amplitude and the quasi two-body nature of the FSIs. Therefore,
the success of our description for $m_{K\pi}\lesssim 1.5$~GeV suggests
that, in this domain, the physics of the decay is dominated by
two-body $K\pi$ interactions. We are led to conclude that effects not
included in our model such as the $I=3/2$ non-resonant $K\pi$ $S$
wave, the non-resonant $I=2$ $\pi^+\pi^+$ interactions and genuine
three-body interactions, could be considered as corrections to the
general picture described here.

Part of the discrepancy observed in our Dalitz plot is due to the
disaccord of our $S$-wave amplitude for $m_{K\pi}\gtrsim
1.5$~GeV. A possible cause for this disagreement is the fact that
factorisation in a three-body decay is expected to break down close to
the edges of the Dalitz plot~\cite{BenoitBKpipi,Beneke}. Furthermore,
in this region, the kinematical configuration of the final state
momenta renders the quasi two-body treatment less trustworthy as
well. Finally, our model does not include the tensor
component. Although marginal, this amplitude has a non-trivial
distribution in the phase space and could induce sizable interference
effects in our plots. In the vector channel, we find puzzling that the
$K^*(1410)$, which gives a sizable contribution for $\tau^-\to K \pi
\nu_{\tau}$~\cite{JPP2,Boito:2008fq}, is hardly seen in experimental
analyses of $\DKpp$.

 In conclusion, since we do not fit the Dalitz plot we think that the
 agreement between the model and the experimental data is
 satisfactory.

%%%%%%%%%%% Acknowledgements %%%%

\section*{Acknowledgements}
We thank M.~Jamin for encouraging us to perform this work, for
providing the tables for the scalar form factor of
Ref.~\cite{Jamin:2006tj}, and for a careful reading of the
manuscript. We also thank A.~A.~Machado, A.~Polosa, and
J.~J.~Sanz-Cillero for useful discussions.  This work has been
supported in part by the Ramon y Cajal program (R.~Escribano), the
{\it Ministerio de Ciencia e Innovaci\'on} under grant
CICYT-FEDER-FPA2008-01430, the EU Contract No.~MRTN-CT-2006-035482
``FLAVIAnet'', the Spanish Consolider-Ingenio 2010 Programme CPAN
(CSD2007-00042), and the {\it Generalitat de Catalunya} under grant
SGR2005-00994. We also thank the {\it Universitat Auton\`oma de Barcelona.}

%%%%% APPENDIX %%%%%%%%%%%%%%%%
\appendix

\section{$K\pi$ form factors}
\label{formfactors}

The scalar and vector $K\pi$ form factors employed in this work were
obtained respectively in Ref.~\cite{Jamin:2006tj}
and Ref.~\cite{Boito:2008fq}. The details can be found in the original
references but for the sake of  completeness we briefly summarise here
how they are obtained.

\subsection{Scalar $K\pi$ form factor}

The framework for the determination of the scalar $K\pi$ form factor,
$F_0^{K \pi}(s)$, is described in detail in Ref.~\cite{Jamin:2001zq}. The
results were numerically updated later and we employed in our
numerical analysis the latest version given in Ref.~\cite{Jamin:2006tj}.
In Ref.~\cite{Jamin:2001zq}, the authors solved a generalised Omn\`es
problem where three channels, namely $K\pi$, $K\eta$ and $K\eta'$,
are taken into account. In this framework, the scalar form factor for channel $k$, $F_0^{k}(s)$ (where $1\equiv K\pi$, $2\equiv K\eta$ and $3\equiv K\eta'$), can be cast as a sum over the three channels as
\beq
F_0^k(s)= \frac{1}{\pi} \sum_{j=1}^3\,  \int\limits_{s_j}^{\infty} d s'\, \frac{\sigma_j(s')F_0^j(s')t_0^{{k\to j}}(s')^*}{(s' -s -i\epsilon)}\ . \label{F0}
\eeq
In the last equation, $s_j$ is the threshold for channel $j$,
$\sigma_j(s)$ are two-body phase-space factors and $t_0^{k\to j}$ are
partial wave $T$-matrix elements for the scattering $k\to j$. The form
factors are obtained solving the coupled dispersion relations arising
from Eq.~(\ref{F0}). This is done imposing chiral symmetry constraints
and using $T$-matrix elements from Ref.~\cite{JOPscatt}  that provide a
good description of scattering data.  One recovers the elastic
approximation by considering solely the contribution of the channel
$k$ to the right-hand side of Eq.~(\ref{F0}), which  is then
reduced to the usual Omn\`es equation~\cite{Omnes:1958hv}.

\subsection{Vector $K\pi$ form factor}

The vector $K\pi$ form factor, $F_+^{K\pi}(s)$, employed in this work was obtained
in Ref.~\cite{Boito:2008fq} within a dispersive representation from fits
to $\tau^- \to K \pi \nu_\tau$ data obtained by the Belle collaboration~\cite{Belle}.
The reduced vector form factor $\tilde F_+^{K\pi}(s)\equiv F_+^{K\pi}(s)/F_+^{K\pi}(0)$ is written
 in terms of a three-times-subtracted dispersion relation
that takes the form
\begin{equation}
\label{KpivffDR}
\tilde F_+^{K\pi}(s)=\exp\left[\alpha_1\frac{s}{m_\pi^2}+\frac{1}{2}\alpha_2\frac{s^2}{m_\pi^4}
+\frac{s^3}{\pi}\int\limits_{s_{K\pi}}^{s_{\rm cut}}
ds^\prime\frac{\delta_1^{K\pi}(s^\prime)}{(s^\prime)^3(s^\prime-s-i0)}\right]\ ,
\end{equation}
where $s_{K\pi}$ is the $K\pi$ threshold and $\delta_1^{K\pi}$ is the form-factor phase.
The subtraction constants $\alpha_1$ and $\alpha_2$
can easily be related to the slope parameters $\lambda_+^{(n)}$,
which appear in the Taylor expansion of $\tilde F_+^{K\pi}(s)$ around $s=0$,
\begin{equation}
\label{KpivffTaylor}
\tilde F_+^{K\pi}(s)=1+\lambda^\prime_+\frac{s}{m_\pi^2}
+\frac{1}{2}\lambda^{\prime\prime}_+\frac{s^2}{m_\pi^4}+\cdots\ ,
\end{equation}
as $\lambda^\prime_+=\alpha_1$ and
$\lambda^{\prime\prime}_+=\alpha_2+\alpha_1^2$.  The cutoff $s_{\rm
  cut}$ is introduced as the upper limit of the Omn\`es integral to
study the importance of the high-energy region which is strongly
suppressed by the factor $s'^3$ in the denominator of the integrand of
Eq.~(\ref{KpivffDR}).  Furthermore, within the elastic region,
$\delta_1^{K\pi}$ is the $P$-wave $I=1/2$ $K\pi$ scattering phase
shift.  An advantage of the three-times-subtracted form of $\tilde
F_+^{K\pi}(s)$ is to make the  results less sensitive to deficiencies
of the phase shift in the higher-energy region.  Then, the integral in
Eq.~(\ref{KpivffDR}) emphasises the lower-energy domain (elastic
domain), for which one can provide a reliable model for the phase
shift.  The description of $\delta_1^{K\pi}$ we used is inspired by RChT and includes the
contribution of two vector resonances namely the $K^*(892)$ and the $K^*(1410)$. The detailed expressions can be found
in Ref.~\cite{Boito:2008fq}.

%%%%%%%%%%%%% BIBLIOGRAPHY %%%%%%%%%%%

\end{document}